\pdfoutput=1

\documentclass[structabstract]{aa}  

\usepackage{graphicx}
\usepackage{txfonts}
\usepackage{natbib}
\usepackage{rotating}
\usepackage{wasysym}

\newcommand{\ergpersec}{erg\,s$^{-1}$}
\newcommand{\ergpercm}{erg\,cm$^{-2}$}
\newcommand{\ergperseccm}{erg\,s$^{-1}$\,cm$^{-2}$}
\newcommand{\gperseccm}{g\,s$^{-1}$\,cm$^{-2}$}
\newcommand{\gpercm}{g\,cm$^{-2}$}
\newcommand{\ergperg}{erg\,g$^{-1}$}

\newcommand{\cm}{cm$^{-2}$}

\newcommand{\sgra}{Sgr\,A$^\star$}

\newcommand{\grs}{GRS\,1741.9--2853}
\newcommand{\grsax}{AX\,J1745.0--2855}
\newcommand{\grsigr}{IGR\,J17453--2853}
\newcommand{\grscxo}{CXOGC\,J174502.3--285450}

\newcommand{\xmm}{\it XMM-Newton\rm}
\newcommand{\integ}{\it INTEGRAL\rm}
\newcommand{\granat}{\it Granat\rm}
\newcommand{\sax}{\it BeppoSAX\rm}
\newcommand{\asca}{\it ASCA\rm}
\newcommand{\chandra}{\it Chandra\rm}
\newcommand{\swift}{\it Swift\rm}
\newcommand{\rxte}{\it RXTE\rm}

\begin{document}

   \title{Bursting behavior of  the Galactic Center \\
   faint X-ray transient \grs}

   \subtitle{}

   \author
   	{G. Trap
          \inst{1,2}\fnmsep\thanks{\email{trap@apc.univ-paris7.fr}}
          \and
          M. Falanga
          \inst{1,3}
          \and
          A. Goldwurm
          \inst{1,2}
          \and
          E. Bozzo
          \inst{4,5}
          \and
          R. Terrier
          \inst{2}
          \and
          \\
          P. Ferrando
          \inst{1,2}
          \and
          D. Porquet
          \inst{6}
          \and 
          N. Grosso
          \inst{6}
          \and
          M. Sakano
          \inst{7}
                 }

   \institute
   	{Service d'Astrophysique (SAp) / IRFU / DSM / CEA Saclay -- B\^at. 709, 91191 Gif-sur-Yvette Cedex, France
         \and
           AstroParticule \& Cosmologie (APC) / Universit\'e Paris VII / CNRS / CEA / Observatoire de Paris -- B\^at. Condorcet, 10, rue Alice Domon et L\'eonie Duquet, 75205 Paris Cedex 13, France
           \and
           International Space Science Institute (ISSI), Hallerstrasse 6, CH-3012 Bern, Switzerland           
           \and
           Istituto Nazionale Di Astrofisica (INAF) / Osservatorio Astronomico di Roma, Via Frascati 33, Monte Porzio Catone, 00044 Rome, Italy
           \and
           INTEGRAL Science Data Center (ISDC) / Science data center for Astrophysics, Ch. d'Ecogia 16, CH-1290 Versoix (Ge), Switzerland
           \and
           Observatoire astronomique de Strasbourg / Universit\'e de Strasbourg / CNRS / INSU -- 11, rue de l'Universit\'e, 67000 Strasbourg, France
           \and
           Department of Physics and Astronomy / University of Leicester -- LE1 7RH Leicester, UK
            }

   \date{Received ... ; accepted ...}
  \abstract
     {}
   {The neutron star low-mass X-ray binary \grs~is a known type-I burster of the Galactic Center. It is transient, faint, and located in a very crowded region, only $10'$ from the supermassive black hole \sgra. Therefore, its bursting behavior has been poorly studied so far. In particular, its persistent emission has rarely been detected between consecutive bursts, due to lack of sensitivity or confusion. This is what made \grs~one of the nine "burst-only sources" identified by \sax~a few years ago. The physical properties of \grs~bursts are yet of great interest since we know very little about the nuclear regimes at stake in low accretion rate bursters. We examine here for the first time several bursts in relation with the persistent emission of the source, using \integ, \xmm, and \swift~observations.}
   {We investigate the source flux variability and bursting behavior during its 2005 and 2007 long outbursts. These events were almost entirely covered by \integ, both in the soft (3--20~keV) and hard (20--100~keV) X-ray bands, and also partly by \xmm~and \swift~(2--10~keV) in 2007.}
   {In its last activity periods, 2005 and 2007, the persistent luminosity of \grs~varied between $\sim$1.7 and $10.5\times10^{36}$~\ergpersec, i.e. $0.9-5.3$\% of the Eddington luminosity. 
   The shape of the spectrum as described by an absorbed power-law remained with a photon index $\Gamma \approx 2$ and a column density $N_{\rm H}\approx 12\times10^{22}$~\cm~throughout the outbursts. We discovered 11 type-I bursts with \integ, and inspected four additional bursts: two recorded by \xmm~and two by \swift. From the brigthest burst, we derive an upper limit on the source distance of $\sim$7~kpc. The observed bursts characteristics and source accretion rate suggest pure helium explosions igniting at column depths $y_{\rm ign}\approx 0.8-4.8 \times 10^{8}$~\gpercm, for typical energy releases of $\sim$$1.2-7.4\times10^{39}$~erg.}   
   {}

   
   \titlerunning{Bursting bahavior of \grs}
\authorrunning{G. Trap et al.}
   \maketitle
 

\section{Introduction}

A low-mass X-ray binary (LMXB) is a stellar system consisting of a low mass star ($<1M_\odot$) filling its Roche lobe and pouring matter onto a compact object via an accretion disk. A large fraction ($\sim$55\%) of the LMXB of our galaxy are transient \cite[see][for a recent catalogue]{liu07}, which means they display bright X-ray outbursts from time to time, lasting from weeks to months. During such an event, they typically increase their luminosity by a factor of at least a hundred compared to the dim state in which they spend most of their time and are usually too faint to be detected. One traditionally distinguishes three classes of transients according to their 2--10~keV peak luminosities: \textit{bright} transients ($10^{37-38}$~\ergpersec), \textit{faint} transients ($10^{36-37}$~\ergpersec), and \textit{very faint} transients ($10^{34-36}$~\ergpersec) \citep[e.g.,][]{wijnands06}. In particular, \textit{faint} transients were recognized as a distinctive class with the observations of \sax~a decade ago \citep{heise99}. 
Just like for other transients, the outbursts of the faint ones can be accomodated by the classical disk instability model \citep[][for a review]{king98,lasota01}, provided their orbital periods are shorter than 80~min \citep{king00}. Yet, these faint transients are markedly  concentrated towards the Galactic Center (GC) \citep{intzand01}, and generally harbor a neutron star as compact object, which is inferred from the presence of type-I X-ray bursts in the light curves of such systems. Indeed, type-I bursts
are thought to result from thermonuclear explosions on the solid surface of the neutron star: see \citet[][]{lewin93,strohmayer06} for reviews, and \citet[][]{cornelisse03,galloway08,chelovekov07} for recent catalogues. 
Interestingly, the properties of the bursts are tightly bound to the accretion rate of the neutron star. Early theoretical work by \cite{fujimoto81} showed that, at low accretion rates, the bursts should be triggered by the unstable burning of accreted hydrogen and should thus be dissimilar to the usual "pure helium" bursts encountered at higher accretion rates. From an observational stand point, the study of low accretion rate bursters developped with the discovery of nine "burst-only sources" by \sax~\citep{cocchi01,cornelisse04}. Indeed, the persistent emission---and so the accretion rate---of these objects was so low that the source could never be detected right before and after the bursts. This subclass of sources also recently motivated theoretical studies suggesting that, at the lowest accretion rates, unstable H burning triggers vigourous He flashes in a H-rich environment, whereas at slightly higher acretion rates, weak pure H flashes build up a thick He layer underneath the surface \citep{peng07,cooper07}. This layer can eventually ignite, leading to a several minutes long burst. Dim bursters could therefore possibly help bridge the gap between short bursts and the growing set of intermediate long bursts brought to light in the past 10 years \citep[][and references therein]{falanga08,falanga09}. We also emphasize that almost all currently known X-ray recycled millisecond pulsars are faint LMXB transients \citep{wijnands06b,falanga08b}, making these sources interesting targets to look for fast rotators.

In this context, we present the observational analysis of one of the "burst-only sources", \grs, with all the data pertaining to the GC ever collected by \integ~and \xmm~since February 2002, and \swift~data from early 2007.    

\subsection{History of \grs~from 1990 to 2004}

 \grs~was discovered in the GC in spring 1990 by ART-P, the low energy instrument onboard the \granat~satellite \citep{sunyaev90,sunyaev91}. The source was seen at a constant flux level of $2 \times 10^{-10}$~\ergperseccm~(4--30~keV) during two observations in March and April, and was also present in the mosaics of \granat~high energy instrument SIGMA \citep[40--100~keV,][]{goldwurm96}. But in all the following campaigns on the GC however, \granat~failed to detect \grs, establishing the source as transient. Note that this binary is only $10'$ from the Milky Way supermassive black hole \sgra, and was thus in the field of view (FOV) of numerous observations carried out by high energy satellites over the years (see below). 
In September of 1994, \asca~identified \grs~(a.k.a. \grsax) back in outburst with a flux of $1.8\times 10^{-12}$~\ergperseccm~in the 0.7--10~keV band \citep{sakano02}. 
 
Two years later (August and September 1996), \sax~recorded three type-I X-ray bursts from \grs~with its Wide Field Camera \citep{cocchi99}, implying that the source is a LMXB housing an accreting neutron star. From a photospheric radius expansion (PRE) episode in the brightest burst, \cite{cocchi99} derived a distance of $\sim$8~kpc for \grs, placing it close to the very center of the galaxy. 
This was based on the assumption that, during PRE, the bolometric peak luminosity of a burst saturates at the Eddington luminosity and thus acts as a standard candle \citep{lewin93,kuulkers03}.
Note that \sax~could not measure the persistent flux of the source and set an upper limit on the bolometric luminosity of $\sim$$1.6\times 10^{36}$~\ergpersec, which made this burster another member of the "burst-only sources" class \citep{cornelisse04}. A little earlier in 1996, from April to July, \rxte~scanned the GC and detected eight bright bursts from an unidentified source in this intricate region. Even though the FOV of \rxte/PCA contained many bursters, \cite{galloway08} attributed the bursts to \grs~in light of the activity seen by \sax~at roughly the same time and the similarity between the bursts caught by the two satellites. Interestingly, three of the \rxte~bursts showed burst oscillations at a frequency of 589~Hz \citep{strohmayer97}, so that if the association of the source and the bursts holds, the neutron star in \grs~is spinning with a period of only 1.7 ms. Due to confusion, the PCA could not measure the persistent flux of the source. 
Unlike \rxte, \asca~managed to detect this persistent emission from \grs~in September 1996 \citep{sakano02}, and so strengthened the association of the \rxte~and \sax~bursts into a long outburst of the source that lasted at least six months.
Thanks to good statistics, physical spectral parameters were constrained. The persistent emission X-ray spectrum was fitted with a photoelectrically absorbed power-law model, whose parameters are the column density, $N_{\rm H} = 11.4^{+0.9}_{-0.8}\times10^{22}$~\cm, and the spectral photon index, $\Gamma = 2.36\pm0.16$. The average flux was $1 \times 10^{-10}$~\ergperseccm~(0.7--10~keV).

The binary was undetected during another four years, until \chandra~witnessed the source in outburst in fall 2000, with an associated weak thermonuclear burst \citep{muno03b}\footnote{\grs~is also known as \grscxo~in the \chandra~catalogue \citep{muno06}.}. These authors fitted the spectrum of the persistent emission with an absorbed power-law, $N_{\rm H} = (9.7\pm 0.2)\times10^{22}$~\cm~and $\Gamma = 1.88\pm 0.04$. They found the source in quiescence in summer 2001 too, at a luminosity of $\sim$$10^{32}$~\ergpersec~(2--8~keV). These \chandra~measurements provided the most accurate position of \grs~to date: $\alpha=17^{\rm h}45^{\rm m}2.33^{\rm s}, \delta=-28^\circ54'49.7''$ (J2000), with an uncertainty of $0.7''$. By browsing the \xmm~archive, we found that in October 2002 \grs~was again weakly active at a luminosity of $\sim$$10^{35}$~\ergpersec~(2--8~keV) (see Sect.~3.2.1.), which is somewhat reminescent of the weak outburst captured by \asca~in 1994. \xmm~upper limits on the burster flux until 2004 are given in Tab.~\ref{history}.

\subsection{The 2005 and 2007 outbursts}

From February to April 2005, the hard X-ray imager \integ/IBIS/ISGRI perceived a new outburst from \grs~\citep{kuulkers07c}. The source was temporally called \grsigr~until it was later recognized as \grs~\citep{kuulkers07b}. \chandra~found the source still in activity in June \citep{wijnands05} and slowly fading one month later \citep{wijnands06}. The July spectrum of the source was similar to the previous \chandra~fit published earlier by \cite{muno03b}: $N_{\rm H} = 10.5^{+4.9}_{-3.7}\times10^{22}$~\cm~and $\Gamma = 1.8^{+1}_{-0.8}$. These observations triggered an optical follow-up in the I band, that could not identify any visible counterpart \citep{laycock05}, in accordance with the high reddening expected from GC sources. As we will explain below (Section~\ref{sec:burstJem}), during this at least 3 months long outburst, we found 4 type-I X-ray bursts with \integ/JEM-X. 

In 2007, renewed activity from \grs~was detected by several observatories. In February, \integ~started a new long exposure on the Galactic Bulge, and clearly caught the source in activity \citep{kuulkers07a,kuulkers07b}. The satellite monitored the outburst all the way to April when the source went back to quiescence. We discovered a total of seven bursts distributed between March and April (see Sect.~\ref{sec:burstJem}). The early detection of the source by \integ~triggered a short pointing with \swift/XRT~which detected a new X-ray burst \citep{wijnands07}. It is likely that another burst of \grs~was also detected by \swift/BAT earlier, in January \citep{Fox07}. 
These two bursts are also inspected in Section~\ref{sec:burstSwift} of the present paper.
From March to April, \swift/XRT followed the source more regularly, and measured the peak of the outburst at a luminosity of $\sim$$2 \times 10^{36}$~\ergpersec~(2--10~keV) \citep{degenaar09}. The latter authors fitted the average spectrum with $N_{\rm H} = 14^{+1}_{-0.9}\times10^{22}$~\cm~and $\Gamma = 2.6\pm0.2$. They also noticed the presence of a short (one week long) and faint ($\sim$$10^{35}$~\ergpersec) outburst of the source in September 2006, that might have been a precursor to the 2007 longer outburst. \chandra~measured the persistent luminosity of \grs~on February 22$^{\rm nd}$: $4\times10^{35}$~\ergpersec~($N_{\rm H} = 9 \pm 0.5 \times10^{22}$~\cm~and $\Gamma = 0.9\pm0.1$, 2--8~keV) \citep{muno07}. On February $27^{\rm th}$, an \xmm/PN observation was carried out in Timing mode to search for millisecond pulsations in the persistent emission and burst oscillations. Yet, neither was a burst recorded nor was any pulsation found \citep{wijnands08}. Finally, \xmm~was pointed once again to the GC at the beginning of April as part of a multiwavelength campaign on the supermassive black hole \sgra~\citep{porquet08,trap08b,dodds09}. This long observation ($\sim$100~ks) coincided with the end of \grs~outburst and contains two type-I bursts \citep{porquet07}; we examine these in greater details in Section~\ref{sec:burstXmm}.

\begin{figure*} 

   \centering
   \includegraphics[trim=0cm 9.5cm 0cm 10cm, clip=true, width=13cm]{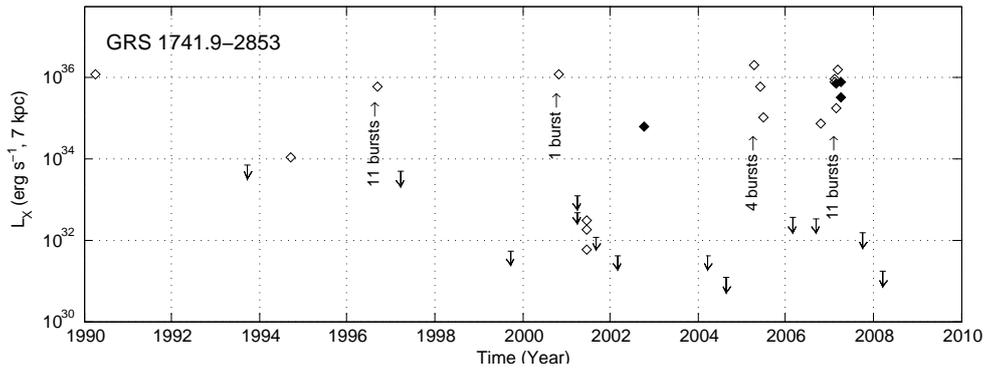}
   \caption{X-ray luminosity history of \grs. Open diamonds and upper limits before 2002 correspond to data published elsewhere, and filled diamonds to the \xmm~measurements reported here. See Table \ref{history} for the respective energy ranges of the different points. We indicate the total number of bursts detected during each outburst by \sax, \rxte, \chandra, \integ, \swift, and \xmm.}
   \label{lc_persist}
\end{figure*}

To sum up \grs~flux history, we have updated in Fig.~\ref{lc_persist} the long term light curves extracted from \cite{muno03b} and \cite{wijnands06}, with data spanning from 2002 to 2007 (Table \ref{history}). Over the past 18 years, the X-ray luminosity of \grs~varied regularly from $10^{32}$ \ergpersec~in quiescence, up to $10^{36}$ \ergpersec~in outburst, justifying the source classification of faint soft X-ray transient.

\begin{table}

\renewcommand{\footnoterule}{}  
\begin{minipage}[t]{\columnwidth}
\caption{\grs~flux history.}
\centering
\begin{tiny}
\label{history}

\begin{tabular}{lccccc}
\hline \hline
Date       & Instrument   & Flux\footnote{X-ray flux in the ranges 2--8~keV for \xmm~and \chandra, 20--60~keV for \integ/ISGRI, 3--10~keV for \integ/JEM-X, and 2--10~keV for \swift/XRT.}   & Ref.\footnote{References: 
[1] \cite{muno03b}, 
[2] This work, 
[3] \cite{kuulkers07c}, 
[4] \cite{kuulkers07b}, 
[5] \cite{wijnands05}, 
[6] \cite{wijnands06},  
[7] \cite{degenaar09}, 
[8] \cite{kuulkers07a}, 
[9] \cite{wijnands07}, 
[10] \cite{muno07}, 
[11] \cite{porquet07}. }   \\
~ & ~ & [$10^{-11}$ \ergperseccm] & ~  \\
\hline

1990--2002 						& \dotfill			& \dots		& [1] \\	
2002 Feb 26						& \xmm/PN		& $<0.0007$	& [2] \\
2002 Oct 3 						& \xmm/PN		& 1		& [2] \\
2004 Mar 28 						& \xmm/PN		& $<0.0007$	& [2] \\
2004 Aug 31						& \xmm/PN		& $<0.0002$	& [2] \\
2005 Apr 10						& \integ/ISGRI		& 31.9	& [3],[4] \\
2005 Jun 5						& \chandra/HRC	& 10		& [5],[6] \\
2005 Jul 1						& \chandra/ACIS	& 1.7		& [5],[6] \\
2006 Feb 27						& \xmm/PN		& $<0.0058$	& [2] \\
2006 Sep 8						& \xmm/PN		& $<0.0057$	& [2] \\
2006 Sep	$\sim$20					& \swift/XRT		& 1.2		& [7] \\
2007 Feb 15						& \integ/ISGRI		& 13		& [4],[8] \\
2007 Feb 16						& \swift/XRT		& 15		& [9],[2] \\
2007 Feb 22						& \chandra/ACIS	& 3		& [10] \\
2007 Feb 27						& \xmm/PN		& 11.6		& [2] \\
2007 Mar						& \swift/XRT		& 26		& [7] \\
2007 Apr 2						& \xmm/PN		& 13.1	& [11],[2] \\
2007 Apr 4						& \xmm/PN		& 5.5 	& [11],[2] \\
2007 Sep 6						& \xmm/PN		& $<0.0026$	& [2] \\
2008 Mar 23						& \xmm/PN		& $<0.0003$	& [2] \\

\hline
\end{tabular}
\end{tiny}
\end{minipage}
\end{table}

\section{Observations and data analysis}

\subsection{INTEGRAL}

The \integ~satellite \citep{winkler03} has been regularly scanning the GC twice a year, in spring and fall, since the beginning of the mission in October 2002. Hereafter, we take into account the data of the Joint European X-ray Monitor (JEM-X), module~1 and 2 \citep{lund03}, between 3 and 20~keV, and the \integ~Soft Gamma-Ray Imager (ISGRI, 20--100~keV) \citep{lebrun03}, from the Imager on Board the \integ~Satellite (IBIS) \citep{ubertini03}. The data were processed with the standard Offline Science Analysis (OSA) software package, version 7.0, distributed by the \integ~Science Data Center \citep{courvoisier03} and based on algorithms described in \cite{goldwurm03b} for IBIS, and \cite{westergaard03} for JEM-X. 

\begin{figure} 
 \centering
  \includegraphics[trim=1cm 1.4cm 15cm 0cm, clip=true, width=7.5cm,angle=0]{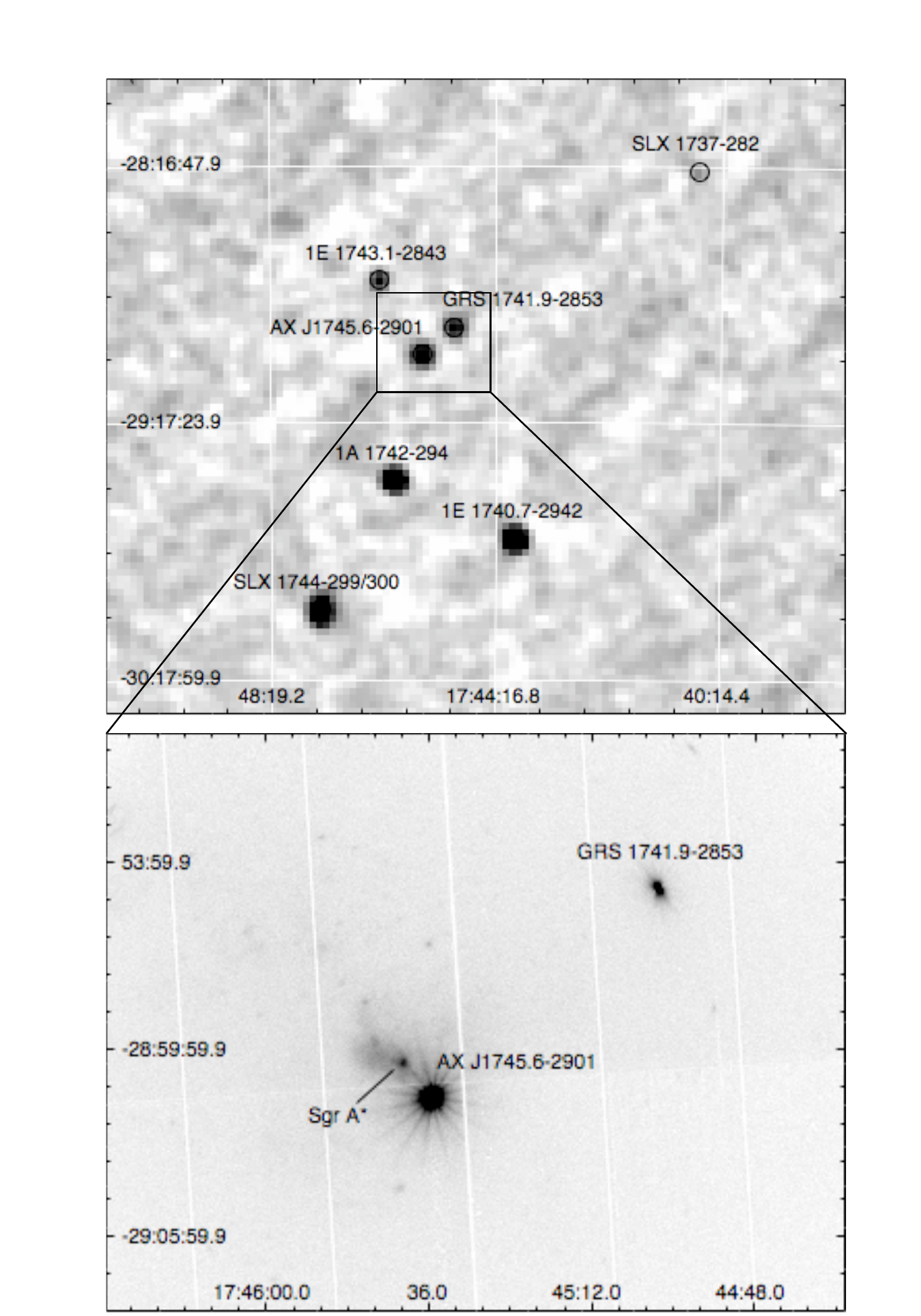}
   \caption{\integ/JEM-X 3--20~keV significance mosaic (top pannel) constructed from 253 individual pointings between February and April 2007, with an effective exposure of 242~ks at the position of \grs. Equatorial coordinates are used and the galactic plane runs from upper left to bottom right. The magnified view (lower panel) is an \xmm/PN 0.1--10~keV image in April 2007 (rev-1340, 97.6~ks exposure). \sgra, the supermassive black hole of the Galactic Center, is clearly apparent since this observation contains several flares from the vicinity of the black hole, enhancing its average luminosity \citep{porquet08}. }
   \label{zoom}
\end{figure}

To search for periods of activity of \grs~with JEM-X, we selected all the public data since the launch of the satellite, pointing less than 3.5$^\circ$ away from the source position. The available data consist of 2626 usable individual pointings. These single pointings were first deconvolved and analyzed separately. We then divided this dataset into six months periods and combined the corresponding individual pointings into mosaics. \grs~was clearly detected only during two periods, from February to April 2005 and February to April 2007, at a significance level of 21$\sigma$ and 12$\sigma$, respectively. Figure \ref{zoom} (upper panel) shows the significance map of the GC region, in the 3--20~keV energy range, for the 2007 dataset. Many sources are present in the FOV of all the mosaics, but thanks to JEM-X high angular resolution \citep[$3.3'$ FWHM at best,][]{brandt03}, \grs~is free of contamination from the neighboring objects. Thereafter, we concentrate on these two epochs, 2005 and 2007, which gather 417 and 253 individual pointings of $\sim$1.8~ks each, for total effective exposures of about 397 ks and 242 ks, respectively.

\subsection{XMM-Newton}

\grs~has been detected on three occasions by the \xmm~satellite \citep{jansen01} operating in different intrument configurations (see Table \ref{log} for a journal of the observations): October 2002, February 2007, and March/April 2007. 
The Science Analysis Software (SAS, version 7.1.2) and the latest Current Calibration files (CCFs) were used to produce calibrated event lists from the Observation Data Files (ODFs). PN event lists were then filtered with the standard selection criteria {\small \tt Flag=0} in order to remove bad pixels, hot pixels, and pixels close to CCD gaps. 

For all the observations, we computed the light curve of the whole detector in the 10--12~keV energy range to identify the time intervals contaminated by background soft proton flares. We excluded from our analysis all the phases where this background count rate exceeded 100 counts per time bins of 100~s. In particular, the last seven hours of both rev-1339 and rev-1340 were rejected in this context.
Apart from rev-1322, where the standard technique for Timing data was applied, all spectra from the persistent emission were extracted from a circular region of $50''$ radius, encircling 90\% of the Point Spread Function (PSF) of the instrument. Background counts were obtained from a similar region offset from the source position. Because the bursts periods were affected by pile-up, special care was taken for the light curves and spectra of these time intervals: events from the inner $6''$ of the PSF were ignored and the parameter {\small \tt Pattern} was set to zero to reject all but single events. For each burst, we took the persistent emission before the burst as background. Version 11.3.2. of the XSPEC software \citep{arnaud96} was used to fit all the spectra with physical models. Errors are quoted at 90$\%$ confidence level for one parameter of interest.

Concerning the observations where \grs~was not active, i.e. February 2002, March/August 2004, February/September 2006, September 2007, and March 2008, we calculated 3$\sigma$ upper limits in counts\,s$^{-1}$ with {\tt \small ximage} and converted them into 2--8~keV unabsorbed fluxes thanks to {\tt \small WebPIMMS}. Results are summarized in Table \ref{history}.

\begin{table}
\begin{minipage}[t]{\columnwidth}
\caption{\xmm/PN and \swift/XRT log of the observations used in this paper.}
\label{log}
\centering
\renewcommand{\footnoterule}{}  
\begin{tiny}
\begin{tabular}{lccccc}
\hline \hline

Orbit		&	ObsID       & 	Date Start    & Exposure & Mode\footnote{The \xmm~optical filter employed is given in brackets: T and M stand for Thick and Medium, respectively.}  \\
~ & ~ & [UTC] &  [ks]  & ~ \\
\hline
                   &                          & \xmm/PN                                                               &              &                   \\
\hline
406		& 111350101	& 2002 Feb 26, 06:41:42				&40.0	& Imaging (T) \\
516		& 111350301	& 2002 Oct 3, 07: 17:07				& 15.3	& Imaging (T) \\
788       	& 202670501	& 2004 Mar 28, 16:45:33				& 41.9	& Imaging (M) \\
866       	& 202670701	& 2004 Aug 31, 03:34:59				& 126.7	& Imaging (M) \\
1139		& 302882601	& 2006 Feb 27, 04:27:37				& 0.5		& Imaging (M) \\
1236		& 302884001	& 2006 Sep 8, 17:19:46				& 0.5		& Imaging (M) \\
1322		& 506291201	& 2007 Feb 27, 06:27:39				& 37.6	& Timing (M) \\
1338		& 402430701	& 2007 Mar 30, 21:28:14		 		& 32.3	& Imaging (M) \\ 
1339		& 402430301	& 2007 Apr 1, 15:07:59				& 103.5	& Imaging (M) \\
1340		& 402430401	& 2007 Apr 3, 16:39:17				& 97.6  	& Imaging (M) \\
1418		& 504940201	& 2007 Sep 6, 10:28:34				& 11.1  	& Imaging (M) \\
1518		& 505670101	& 2008 Mar 23, 17:22:10				& 96.6  	& Imaging (M) \\
\hline
                   &                          & \swift/XRT                                                               &              &                   \\
\hline
\dotfill	& 0003088801	& 2007 Feb 16, 21:39:18				& 3.9  	& Photon counting \\
\hline
\end{tabular}
\end{tiny}
\end{minipage}
\end{table}

\subsection{Swift}

For the present study we used the publicly accessible observations id.~00257213000 and 00030888001 from \swift~\citep{gehrels04}. The first dataset
includes the burst discovered on January 22$^{\rm nd}$, 2007 with the high energy coded mask detector BAT
\citep{Fox07}, whereas the second
comprises both the burst and the persistent emission
observed with the low energy telescope XRT on February 16--17$^{\rm th}$, 2007 \citep{wijnands07}.
After the burst detected by
the \swift/BAT on January 22$^{\rm nd}$, no XRT fullow-up has been performed.
The total exposure time was $\sim$4~ks for BAT during
observation 00257213000, and $\sim$3.9~ks for XRT during
observation 00030888001.

For the BAT data analysis, we used the {\small \tt batgrbproduct} tool
included in the Heasoft package
6.6.1. The BAT spectrum has been extracted over the entire duration of
the burst (9 s). The XRT data were processed with the {\small \tt xrtpipeline}
(version 0.12.1) task. Filtering and screening criteria were applied by using the
{\small \tt ftools} (Heasoft version 6.6.1). We analyzed data in photon counting (PC)
mode, and selected event grades of 0--12. All the XRT/PC data were
affected by a strong pile-up, and thus we extracted the source
light curves and spectra by using annular regions centered on the source.
The persistent light curves and spectra were extracted through an
annular region with an inner radius of 4 pixels and an external radius
of 20 pixels. Unfortunately, the count rate during the
burst was very high and the corresponding pile-up was much stronger than
that of the persistent emission. In order to correct for this effect, we
used annular regions with different inner radii during the rise, the
peak and the decay of the burst (the inner radii were of 10, 15 and
4 pixels, respectively).
We then used the {\small \tt xrtlccorr} task to account for these corrections
in the background-subtracted light curves.
To estimate the spectral properties of the source and its flux during
the entire burst, we used an extraction region with an inner radius
of 10 pixels. All the spectra were rebinned in order to have at least 20
photons per bin so as to permit $\chi^2$ fitting.
Exposure maps were created through the {\small xrtexpomap} routine, and we
used the latest spectral redistribution matrices in the HEASARC
calibration database (version 0.1.1).
Ancillary response files, accounting for different extraction regions,
vignetting and PSF corrections, were generated via
the {\small \tt xrtmkarf} task.

\section{Results}

\subsection{INTEGRAL}
\subsubsection{Outbursts}

\begin{figure}
  \centering
  \includegraphics[trim=3cm 7.5cm 3cm 7.5cm, clip=true, width=9cm,angle=0]{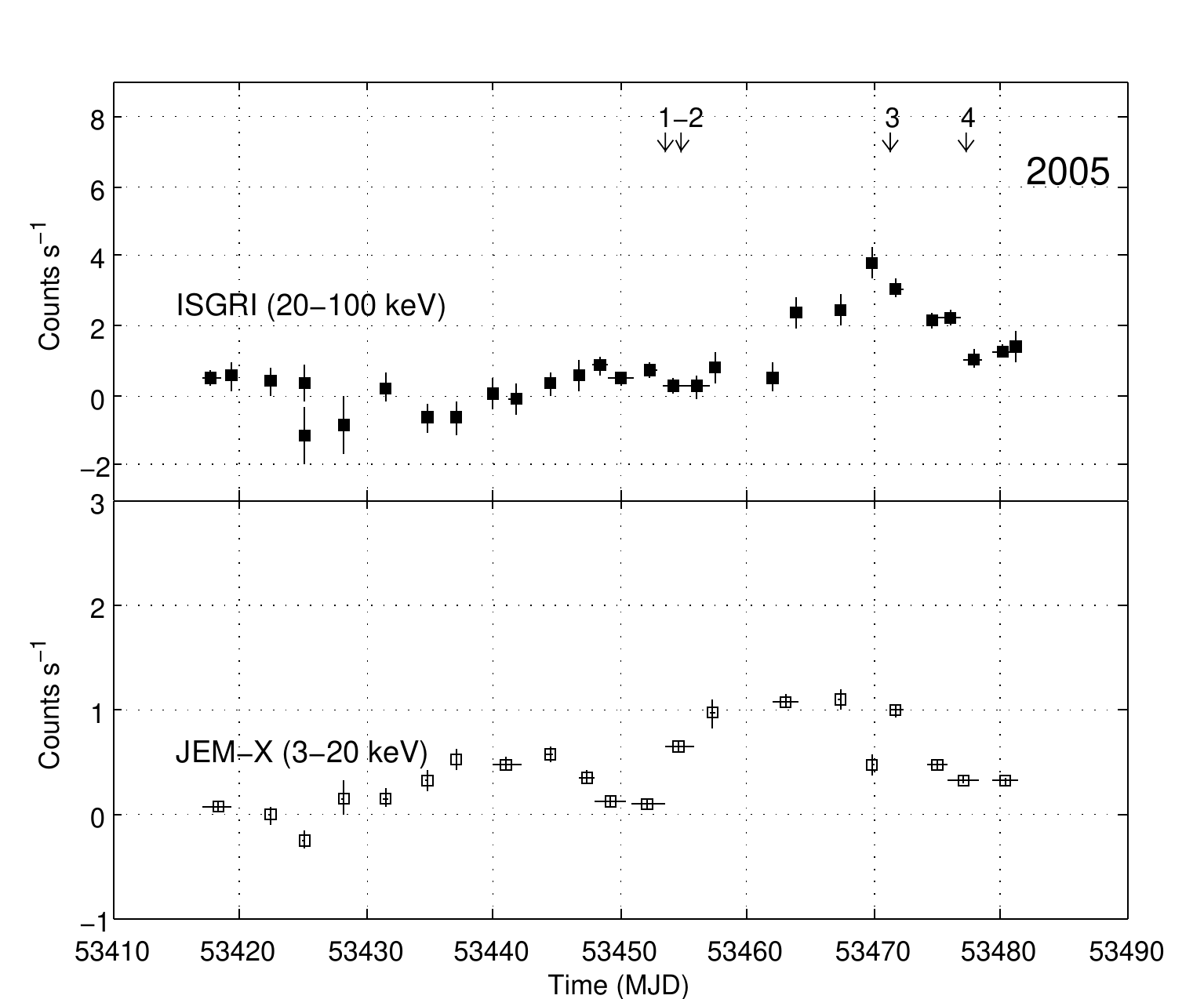}
  \includegraphics[trim=3cm 7.5cm 3cm 7.5cm, clip=true, width=9cm,angle=0]{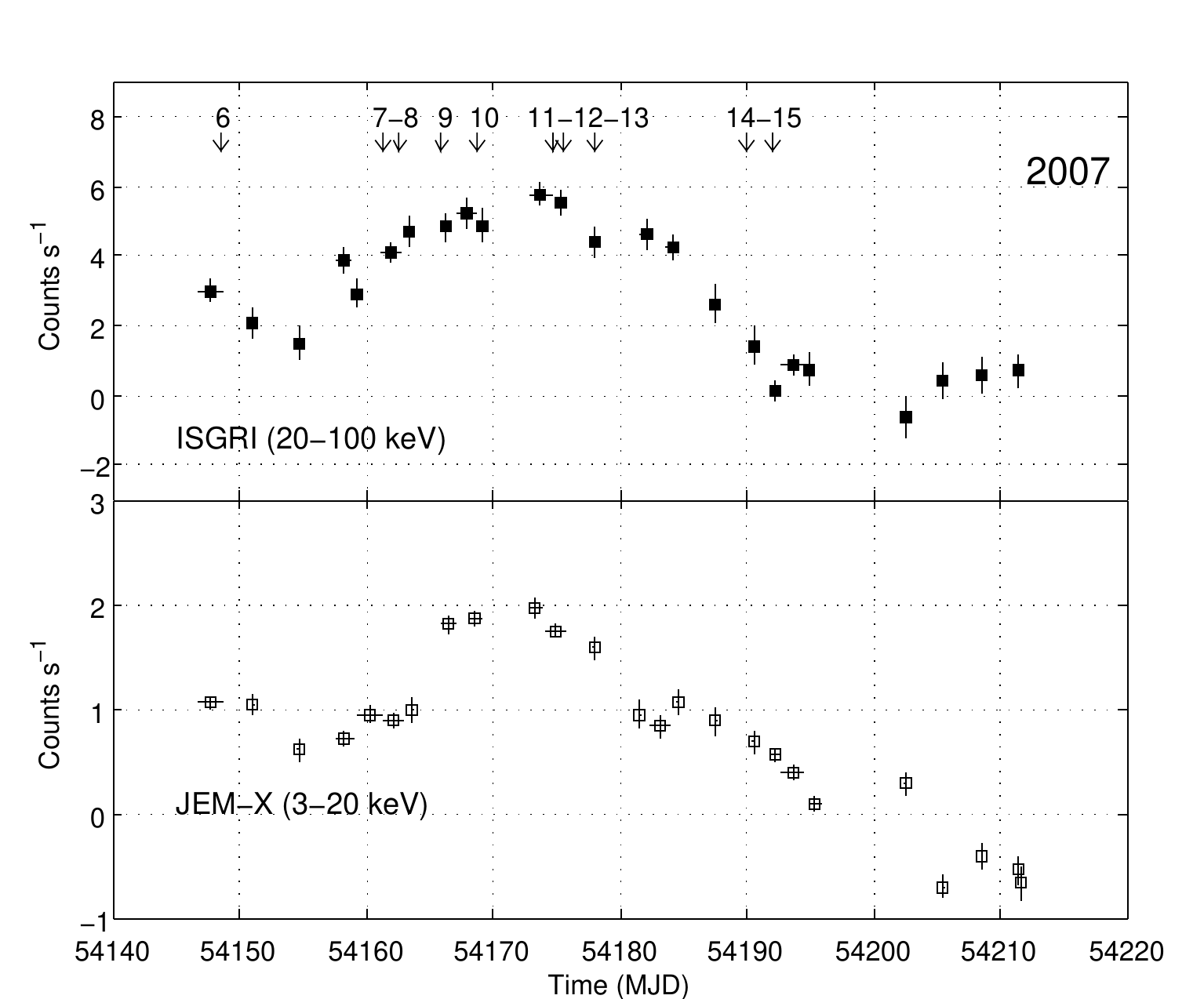}
   \caption{\integ/ISGRI and JEM-X light curves of the 2005 (upper panel) and 2007 (lower panel) outbursts in the 20--100~keV and 3--20~keV bands, respectively. All the type-I bursts reported in this work are indicated by the arrows, leaving out the \swift/BAT burst \#5 that occured on $\sim$MJD~54122. Note that the error bars on the count rate and the binning times are sometime within the square symbols. Negative count rates are due to statistical fluctuations at low flux level after background subtraction.}
   \label{lc_all}
\end{figure}

Albeit significant in the global mosaics of the 2005 and 2007 outbursts, the source was so weak that it was never significantly detected in neither JEM-X nor ISGRI individual pointings. Hence, the standard technique to derive light curves with OSA was not adequate. So, we built the light curves from the individual images, by collecting the count rate in the pixel consistent with the source position for ISGRI, and by integrating the flux over a 4.2$'$ (FWHM) PSF for JEM-X\footnote{The size of the PSF was empirically derived from a bright source of the FOV, GX\,3+1, in the 3--20~keV range.}. Then, by rebinning with long intervals ($\sim$days), the evolution of the outbursts becomes visible (see Fig.~\ref{lc_all}). In 2005, \integ~observations ceased in April, but \chandra~still detected the source fading in June and July \citep{wijnands05,wijnands06} so \integ~missed the end of the ouburst.
In 2007, instead, we missed the beginning and got the end. Notice that the drop in the light curve around MJD 54155 (end of April 2007) was confirmed by independent flux measurements of \swift, \chandra~and \xmm~at lower energies, in the $\sim$2--10~keV band (see Table~\ref{history}).

\subsubsection{Type-I bursts}
\label{sec:burstJem}

To search for type-I bursts, we investigated all the JEM-X, module 1 and 2, data publicly available since the beginning of the mission, pointing less than 3.5$^\circ$ away from the source position, and forced the standard pipeline to construct 2~s rebinned light curves. Since the persistent emission of \grs~was never detected in single pointings, these light curves contain mainly background emission and "spikes". Most of these spikes are due to instrumental artefacts and some to contamination by bright type-I bursts from other sources in the FOV. To identify the bursts origins, we systematically built images of the sky between the beginning and the end of the bursts candidates. When a thermonuclear burst from \grs~occurs, one should locate a significant excess at the position of the source in the image. By this means, we found a total of four and seven type-I bursts from \grs~in the 2005 and 2007 outbursts, respectively, and none during the quiescent state of the source.

We looked for the high energy (18--40~keV) tails of these 11 bursts in the ISGRI data as well. Therefore, we extracted the off-axis corrected light curves with events
selected according to the detector illumination pattern, using a pixel illumination fraction threshold of 0.4. Only the brightest burst, \#4, could be convincingly detected in the light curve and image. In so far as the FOV from IBIS is larger than the one of JEM-X, the total IBIS exposure on \grs~is longer and so it was worth searching for additional bursts in the IBIS data stream. 
To do that, we used the {\it INTEGRAL} Burst Alert System (IBAS) software running at ISDC and dedicated to the real time localization and discovery of gamma-ray bursts, transient X-ray sources and bursts in the IBIS/ISGRI data \citep{mereghetti03}. IBAS found only burst \#4.

\begin{figure} 
   \centering
   \includegraphics[trim=0.5cm 1cm 1.6cm 1cm, clip=true, width=9.cm,angle=0]{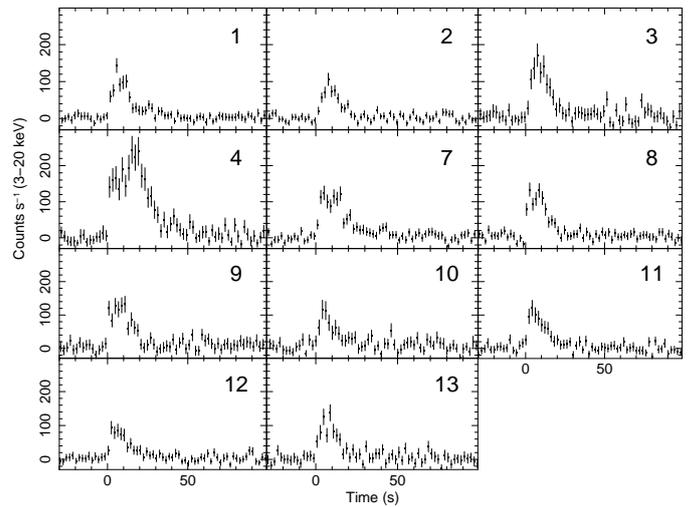}
   \caption{Background subtracted light curves of all the bursts detected by \integ/JEM-X in the 3--20~keV energy range. The time bin is 2 s and the start times of the bursts are given in Table \ref{burstTable}.}
   \label{lc_all_jemx}
\end{figure}

In Table \ref{burstTable}, we report the bursts start times measured by JEM-X. The start time of each burst is defined as the time when the intensity rose to 10$\%$ of the peak above the persistent intensity level.  On Figure \ref{lc_all_jemx}, we plot the background subtracted light curves of the 11 bursts with a 2~s time bin. The background subtraction consisted in removing the constant level present before and after the bursts in the light curve of the individual pointing of interest. Note that this background was not due to the persistent emission of the source, because it was too weak to be detected in such a short period ($\sim$1.8~s). 
The rise time is the interval between the start of the burst and the time at which the intensity reaches 90$\%$ of the peak intensity. For all the bursts it was $\sim$2--4~s, except maybe for burst 4 which had an unusual morphology. The e-folding decay times, determined over the time after the peaks and plateaus, are specified in Table \ref{burstTable}. They roughly range between 10 and 20~s. The total durations, .i.e. from the burst start time back to the persistent flux level in the 3--20~keV band, were around 20--30 s.

For the spectral analysis, we used JEM-X counts during 5~s at the peak of the bursts, in the 3--20~keV band, . The net peak burst spectra are well fitted by a photoelectrically absorbed black-body model ({\small \tt BB}). Unfortunately the energy range covered by JEM-X does not allow us to contrain the interstellar hydrogen column density, $N_{\rm H}$. We therefore fixed $N_{\rm H}$ to $12\times10^{22}$~\cm, the value found with \xmm~(see Sect.~3.2.1.), in all our spectral fits. The inferred {\small \tt BB} temperatures, $kT_{\rm peak}$, and apparent {\small \tt BB} radii, $R_{\rm peak}$, at 7~kpc (see section \ref{sec:discussion}) are listed in Table \ref{burstTable}, along with other bursts parameters. The peak fluxes, $F_{\rm peak}$, were derived from the 3--20~keV light curves peak count  rates with 2~s time resolution and renormalized for the 0.1--100~keV bolometric energy range. The bursts fluences are obtained from the fluxes extrapolated in the bolometric energy band over the respective bursts durations.

To investigate the bursting behavior of a LMXB, it is important to know its accretion rate before the bursts 
(see section \ref{sec:discussion}). A common estimator of this parameter is the persistent luminosity of the source. Here, to assess the persistent flux before one burst, we first built JEM-X mosaics made of between 10 and 30 consecutive pointings preceding the burst in a single band (3--20~keV). Except for bursts 1 and 2 where only upper limits could be set, the persistent emission of the source was always significantly detected in the mosaics. We then converted the count rate fitted in the mosaics into a bolometric flux\footnote{This count rate was compatible with the rebinned light curves on Fig.~\ref{lc_all} but more precise.}, assuming one constant spectral shape throughout the outbursts, composed of an intrinsic power-law of index $\Gamma=2$ between 0.1 and 100~keV (see Sect.~3.2.1.).

\subsection{XMM-Newton}
\subsubsection{Persistent emission}

\begin{figure} 

   \centering
   \includegraphics[trim=3cm 10cm 3.5cm 10cm, clip=true, width=9cm]{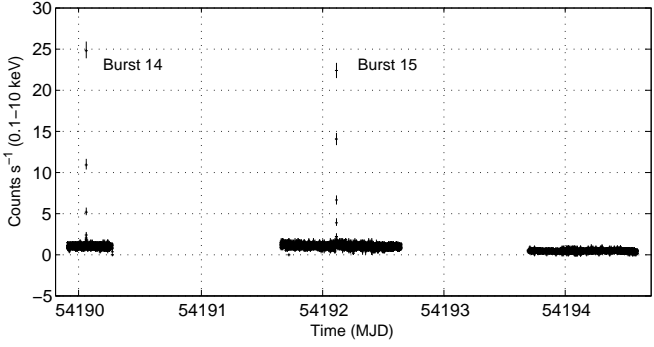}
   \caption{\xmm/PN background subtracted light curve of \grs~in spring 2007 (rev-1338, 1339, and 1340). Two type-I bursts, 14 and 15, are clearly visible.}
   \label{lc_pn}
\end{figure}

The EPIC/PN spectra of \grs~persistent emission in 2002 and 2007 were best fitted by an absorbed power-law. They were fitted between 2--8~keV since, outside these bounds, there were systematic residuals due to the background. Table \ref{specpersist} presents the results of the best fits. We studied all the spectra with $N_{\rm H}$ as a free parameter. Given that we do not notice any apparent variation of $N_{\rm H}$, it is most likely mainly due the galactic absorption. So, we refitted all the data consistently with $N_{\rm H}$ fixed to $11.9\times 10^{22}$~\cm, the value with the smallest error found with the February 2007 pointing.
In 2007, the index remained always close to 2, thus justifying the assumption we made  in Sect.~3.1.2. for the persistent emission measured by JEM-X. The flux was divided by two during last orbit. For this observation, we created an image cleaned for out of time events (OoT) in the 0.1--10~keV, displayed on Fig.~\ref{zoom}. The light curve of the three last observations in March/April 2007 are plotted in Figure \ref{lc_pn}. 

\begin{table}

\begin{minipage}[t]{\columnwidth}
\caption{\xmm/PN and \swift/XRT spectral results for the persistent emission.}
\label{specpersist}
\centering
\renewcommand{\footnoterule}{}  
\begin{tiny}
\begin{tabular}{lcccc}
\hline \hline
Date\footnote{For 2007, Mar 30 and 31 designate the periods just before and after burst 12 respectively (see Fig.~\ref{lc_pn}). Same thing for Apr 1 and 2 and burst 13. Apr 3--4 period is the last \xmm~revolution (rev-1340).}	& $N_{\rm H}$\footnote{Whenever there are brackets, the column density is frozen.} & $\Gamma$ & $F_{\rm pers}$\footnote{Unabsorbed flux in the 2--8~keV and 2--10~keV band in units of 10$^{-11}$ \ergperseccm~for \xmm/PN and \swift/XRT, respectively.} & $\chi^2/{\rm d.o.f.}$ \\
~ & [10$^{22}$ cm$^{-2}$] & ~ & ~  & ~ \\
\hline
                   &                          & \xmm/PN                                                               &              &                   \\
\hline
2002 Oct 3	& $14\pm3$		& $2.0\pm0.5$		& $1.0\pm0.2$ 		& $67/64$\\
~			& $[11.9]$			& $1.6\pm0.2$		& $0.9\pm0.2$ 		& $68/65$\\
2007 Feb 27	& $11.9\pm0.2$	& $1.83\pm0.03$	& $11.60\pm0.05$ 	& $219/197$\\
2007 Mar 30	& $11.6\pm0.4$	& $1.76\pm0.08$	& $11.9\pm0.1$ 	& $155/200$\\
~			& $[11.9]$			& 1.81$\pm0.04$	& $12.2\pm0.1$ 	& $156/201$\\
2007 Mar 31	& $12.1\pm0.4$	& $1.86\pm0.08$	& $12.0\pm0.1$ 	& $192/210$\\
~			& $[11.9]$			& $1.83\pm0.04$	& $11.9\pm0.1$ 	& $192/211$\\
2007 Apr 1	& $12.0\pm0.3$	& $2.08\pm0.05$	& $13.6\pm0.1$ 	& $242/213$\\
~			& $[11.9]$			& $2.06\pm0.02$	& $13.5\pm0.1$ 	& $242/214$\\
2007 Apr 2	& $12.5\pm0.3$	& $2.14\pm0.05$	& $13.1\pm0.1$ 	& $213/195$\\
~			& $[11.9]$			& $2.04\pm0.02$	& $12.4\pm0.1$ 	& $225/196$\\
2007 Apr 3--4	& $13.1\pm0.4$	& $2.30\pm0.08$	& $5.5\pm0.1$ 		& $179/200$\\
~			& $[11.9]$			& $2.08\pm0.03$	& $4.9\pm0.1$ 		& $204/201$\\
\hline
                   &                          & \swift/XRT                                                               &              &                   \\
\hline
2007 Feb 16--17 & $10.9\pm3.6$	& $1.9\pm0.7$		& $19.0\pm3.0$ 		& $31/28$\\
\hline
\end{tabular}
\end{tiny}
\end{minipage}
\end{table}

\subsubsection{Type-I bursts}
\label{sec:burstXmm}

\begin{figure} 
   \centering
 \includegraphics[trim=1.5cm 0.7cm 8.1cm 1.cm, clip=true, width=9.5cm,angle=0]{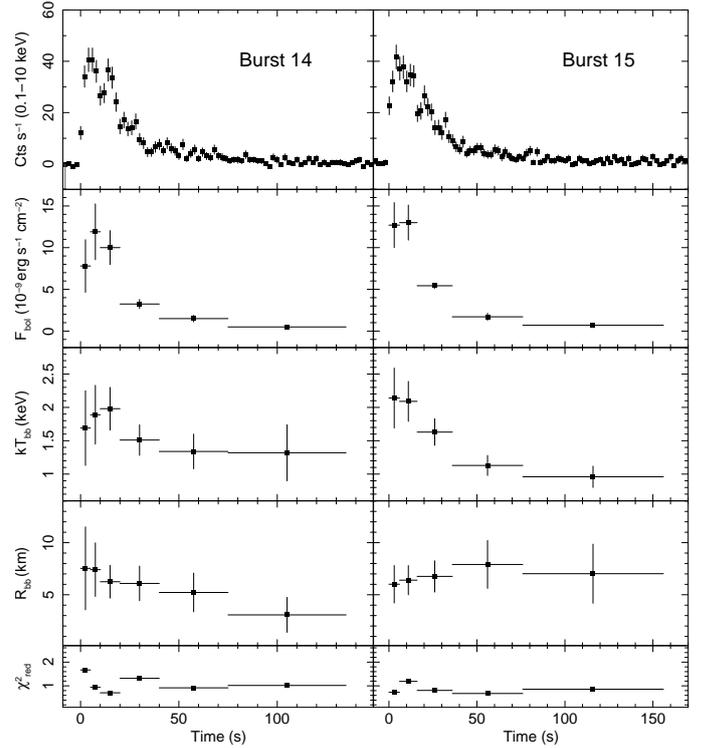}
\caption{Light curves and spectral evolution of bursts 14 and 15 as measured by \xmm/PN in spring 2007. From top to bottom: count rate with a time bin of 2~s, bolometric flux estimated over 0.1--100~keV,  temperature and radius of the photosphere obtained by fitting the spectra with the {\tt \small BB} model, and reduced chi-squared of the fits.}
   \label{specAB}
\end{figure}

In the light curve of the three last revolutions (Fig.~\ref{lc_pn}), we clearly recognize a couple of type-I X-ray bursts (14 and 15) separated by $\sim$178~ks. In view of the huge data gap between them, we cannot claim the bursts were consecutive for sure. A closer view of their net light curves (from which the persistent emission was subtracted) with a time bin of 2~s is visible on Fig.~\ref{specAB} (top pannels). Both bursts have quick rises ($\sim$2~s) followed by exponential decays of about 20~s (see Table \ref{burstTable}). 

The statistics obtained with \xmm~allowed us to perform time-resolved spectral analysis of the bursts. Again, the spectra were well fitted by absorbed black-bodies with $N_{\rm H}$ fixed to $12\times 10^{22}$~\cm. Figure \ref{specAB} shows the evolution of the bolometric flux (0.1--100~keV), $F_{\rm bol}$, black-body temperature, $kT_{\rm bb}$, and black-body radius, $R_{\rm bb}$, as a function of time. In Table \ref{burstTable}, we report these latter values measured at the peak, as well as the fluence, $f_{\rm b}$, integrated over the total duration of the bursts. The first event, \#14, was not contemporaneous with the \integ~GC survey, whereas \#15 fell in an \integ~observing window. Nonetheless we did not detect it neither with ISGRI  nor JEM-X. The ISGRI non detection is consistent wih the fact that the burst was faint, while for JEM-X the absence of detection is surely due to the position of the source in the noisy edge of the FOV.

Regarding for the persistent flux prior to the bursts, we extrapolated the results indicated in Table \ref{specpersist} over the 0.1--100~keV energy band.

\subsection{Swift}
\subsubsection{Persistent emission}

The \swift/XRT persistent emission spectrum of February 16--17$^{\rm th}$ 2007 was fitted by an absorbed power-law model, as used for the \xmm~spectra. The best fit parameters are $N_{\rm H}=10.9\pm3.6 \times10^{22}$~\cm~with a power-law photon index of $\Gamma=1.9\pm0.7$ with a $\chi^{2}_{\rm red}=31.5/28$. The averaged absorbed and unabsorbed 2--10 keV fluxes were $9.8\pm2.5\times10^{-11}$~\ergperseccm~and $19.0\pm0.7\times10^{-11}$~\ergperseccm, respectively. These spectral parameters are in accordance with the previously reported values \citep{wijnands07}. 
Note that this observation was contemporaneous with the beginning of the \integ/JEM-X/ISGRI 2007 observations. The XRT unabsorbed bolometric flux 0.1--100 keV, $\sim$$8.1\times10^{-10}$~\ergperseccm, is comparable with the first pre-burst JEM-X flux reported in Table \ref{burstTable} for burst 7, which occured at the same flux level (see light curve on Fig.~\ref{lc_all} bottom pannels). 
For a long term XRT light curve of \grs, we refer the reader to Fig.~2 in \cite{degenaar09}.

\subsubsection{Type-I bursts}
\label{sec:burstSwift}

\begin{figure} 
   \centering
   \includegraphics[trim=0.5cm 2.5cm 5.5cm 3.5cm, clip=true, width=6.4cm,angle=-90]{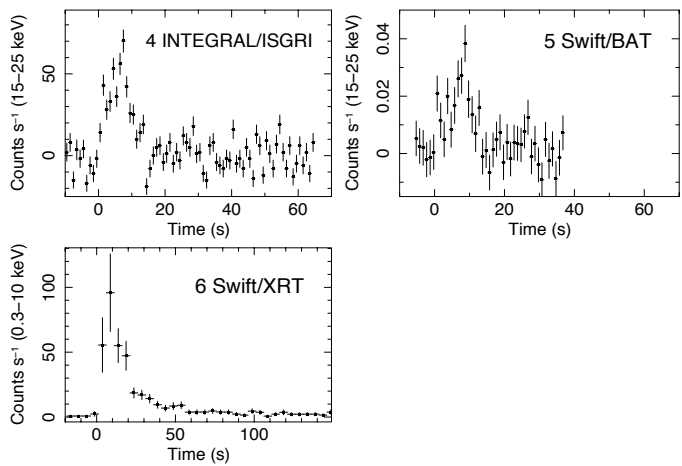}
   \caption{Background subtracted light curves of the \swift~bursts and the ISGRI one for comparison with the BAT one. The time bins are 1~s for \#4 and \#5, and 5~s for \#6. The start times are given in Table \ref{burstTable}.}
   \label{lc_all_swift}
\end{figure}

\swift~detected two type-I bursts: one, \#5, with BAT (15--25 keV) on January 22$^{\rm nd}$, 2007 at 06:12:54 (UTC) and another one, \#6, with XRT 25.69976 days later. For the BAT burst, we cannot determine the cooling time, $\tau$, since only the hard tail of the burst was observed. However, the rise time was around  2~s and the total burst elapsed time was 8.2~s. Due to the short exposure time, the persistent emission could not be measured, in addition, no low energy \swift/XRT follow-up was performed to investigate the persistent emission. The BAT and XRT light curve properties are reported in Table \ref{burstTable}. Notice that the other burst also found at high energy was burst number 4 with \integ/ISGRI. In Fig.~\ref{lc_all_swift} we show the \swift/BAT and \swift/XRT bursts and, for comparison with BAT, the 15--25~keV ISGRI burst.

The BAT burst was not covered by the \integ~scan of the GC and the XRT burst unluckily occured during a slew of \integ, preventing the use of the data.

The spectral analysis of burst \#5 was carried out in the 15--25~keV. The spectrum was well fitted by an absorbed black-body model with the column density again frozen to $12\times10^{22}$ cm$^{-2}$. The inferred bursts parameters are reported in Table \ref{burstTable}. The measured unabsorbed flux was extrapolated to the 0.1--100~keV band by generating {\tt \small dummy} responses with XSPEC. 
such an extrapolation was well justified for the JEM-X data since the black-body temperature was well inside the instrument bandpass (Sec. \ref{sec:burstJem}). However, this is not the case for the BAT burst, so one may question the value of  $kT_{\rm bb}=2.4^{+0.8}_{-0.5}$~keV that we obtain by a simple extrapolation. To estimate the uncertainty introduced by this method, we focused on burst \#4 and compared the fit of the ISGRI data alone and the joint JEM-X/ISGRI spectrum (3--35~keV). We find that fitting solely the high energy spectrum and extrapolating it to softer energies, leads to an error of 30\%, which is inside our BAT error box and so is an acceptable method.   

Due to the high count rate at the peak of burst \#6, the XRT/PC image
was affected by strong pile-up and a spectral analysis of the burst
peak was impossible. Therefore, in order to estimate the burst peak
flux, we used the parameters determined for the burst mean spectrum
($N_{\rm H}$ fixed at $12\times10^{22}$~\cm, and BB temperature
1.6$^{+0.4}_{-0.3}$~keV) and calculated the flux at the peak by using
the burst peak count rate ($96\pm30$~counts\,s$^{-1}$) and the on-line tool {\small \tt webpimms}. We found an unabsorbed 0.1--100~keV flux of $2.7\pm0.8\times10^{-8}$~\ergperseccm.
The burst peak BB temperature could however be higher than the used 1.6~keV.
Therefore, using {\small \tt webpimms} we also estimated the peak flux
for a BB temperature of 3~keV. Since this is believed to be an upper
limit to the X-ray temperature at the burst peak, the corresponding X-ray flux we derived ($6\times10^{-8}$~\ergperseccm) could be reasonably considered an upper limit on the peak X-ray flux as well.

\begin{table*}
\begin{minipage}[t]{\textwidth}
\caption{Bursts parameters.} 
\label{burstTable}
\centering
\renewcommand{\footnoterule}{}  
\begin{tiny}
\begin{tabular}{lccccccccccccc}
\hline \hline
Id.\footnote{Number of the burst, and instrument which detected it. * designates the burst that was also detected by ISGRI.} 
& Burst Start Time 
& $kT_{\rm peak}$\footnote{Black-body temperature at the peak of the burst, assuming $N_{\rm H}=12 \times 10^{22}$~cm$^{-2}$.} 
& $R_{\rm peak}$\footnote{Black-body radius at the peak assuming a distance of 7~kpc to the source.} 
& $\chi^2/$d.o.f. 
& $F_{\rm peak}$\footnote{Unabsorbed bolometric flux (0.1--100~keV) at the peak in units of $10^{-8}$ \ergperseccm.} 
& $f_{\rm b}$\footnote{Fluence of the burst in units of $10^{-7}$ \ergpercm.} 
& $\tau_{\rm fit}$\footnote{E-folding decay time fitted on the light curve.} 
& $\tau_{\rm calc}$\footnote{E-folding decay time calculated with $\tau_{\rm calc}\equiv f_{\rm b}/F_{\rm peak}$.} 
& $F_{\rm pers}$\footnote{Unabsorbed bolometric persistent flux (0.1--100~keV) prior to the burst in units of $10^{-9}$ \ergperseccm.}
& $\Delta t$\footnote{Time ellapsed since last detected burst.} 
& $\gamma$\footnote{Ratio of persistent flux to burst flux: $\gamma \equiv F_{\rm pers}/F_{\rm peak}$.}\\
~ & [UTC] & [keV] & [km] & ~ & ~ & ~ & [s] & [s] &~ & [s] & [$10^{-3}$] \\
\hline
1~~~~JEM-X  & 2005 Mar 24,  14:26:23	& $2.3^{+0.8}_{-0.5}$ & $7^{+5}_{-3}$ 	& 15.7/12 		& $3.2^{+0.7}_{-1.5}$ & 3.6$\pm$0.4& $12^{+2}_{-2}$ 	& 11$\pm$3 & $<0.3$ & \dotfill 			& $<9$ 		\\
2~~~~JEM-X   & 2005 Mar 25,  22:21:43	& $2.0^{+0.5}_{-0.4}$ & $8^{+5}_{-3}$ 	& 8.7/9 		& $2.8^{+0.6}_{-1.6}$ & 2.6$\pm$0.3 & $8^{+2}_{-1}$ 	& 9$\pm$2 & $<0.3$ & 114869 		& $<11$ 		 \\
3~~~~JEM-X   & 2005 Apr 11,  08:53:16	& $1.9^{+0.7}_{-0.5}$ & $12^{+12}_{-12}$ & 13.7/17 	& $4.2^{+1.1}_{-3.4}$ & 4.5$\pm$0.5 & $8^{+2}_{-1}$ 	& 11$\pm$3 & 0.7$\pm$0.2 & 1420269 	& 17$\pm$7 	 \\
4~~~~JEM-X*   & 2005 Apr 17,  06:18:54	& $2.5^{+0.5}_{-0.4}$ & $8^{+3}_{-2}$ 	& 18.4/22 		& $5.5^{+0.7}_{-2.1}$ & 12.6$\pm$1.0 & $11^{+2}_{-2}$ 	& 23$\pm$3 & 0.4$\pm$0.2 & 509114	& 7.7$\pm$4 	 \\
5~~~~BAT   & 2007 Jan 22,  06:12:54 	& $2.4^{+0.8}_{-0.5}$ & $8^{+5}_{-4}$ 	& 5.2/7 		& $5.8^{+1.7}_{-2.5}$ & $>4.7$                 & \dotfill 	&$ >8$ & \dotfill & \dotfill 	& \dotfill 	 \\
6~~~~XRT   & 2007 Feb 16,  22:59:28 	& \dotfill			  & \dotfill		 	& \dotfill 		& $2.7^{+0.8}_{-0.8}$ 			  & 3.3$\pm$0.6 		  & $13^{+6}_{-5}$         & 12$\pm$6 		& 0.8$\pm$0.1 & 	2220459& 29$\pm$12 	 \\
7~~~~JEM-X   & 2007 Mar 02,  08:42:34 	& $1.6^{+0.3}_{-0.3}$ & $15^{+8}_{-5}$ 	& 9.1/13 		& $3.3^{+0.4}_{-1.5}$ & 5.3$\pm$0.5 & $16^{+2}_{-3}$ 	& 16$\pm$3 & 0.7$\pm$0.2 & 1158185 	& 20$\pm$6 	 \\
8~~~~JEM-X   & 2007 Mar 03,  15:51:05 	& $1.9^{+0.3}_{-0.3}$ & $11^{+4}_{-3}$ 	& 11.6/12 		& $3.5^{+0.5}_{-1.2}$ & 3.9$\pm$0.4 & $11^{+1}_{-2}$     & 11$\pm$2 & 1.0$\pm$0.2 & 112057 	& 29$\pm$7 	 \\
9~~~~JEM-X   & 2007 Mar 06,  22:35:16 	& $1.6^{+0.4}_{-0.3}$ & $16^{+13}_{-7}$ 	& 9.1/11 		& $3.6^{+0.6}_{-2.3}$ & 4.3$\pm$0.5 & $10^{+3}_{-2}$ 	& 12$\pm$2 & 1.3$\pm$0.2 & 283423 	& 36$\pm$8 	 \\ 
10~~JEM-X   & 2007 Mar 09,  17:28:56 	& $2.8^{+3.1}_{-1.0}$ & $4^{+6}_{-4}$ 	& 5.9/9 		& $2.8^{+1.2}_{-2.5}$ & 2.5$\pm$0.5 & $12^{+4}_{-2}$ 	& 9$\pm$4 & 1.2$\pm$0.2 & 240790 	& 43$\pm$20 	 \\
11~~JEM-X   & 2007 Mar 15,  16:01:05	& $1.7^{+0.5}_{-0.4}$ & $14^{+12}_{-14}$ & 8.1/10 		& $3.1^{+0.8}_{-2.3}$ & 2.5$\pm$0.4 & $11^{+2}_{-2}$ 	& 8$\pm$3 & 1.8$\pm$0.2 & 513094 	& 58$\pm$16 	 \\
12~~JEM-X   & 2007 Mar 16,  14:09:14	& $2.2^{+1.0}_{-0.6}$ & $7^{+6}_{-7}$ 	& 5.3/8 		& $2.2^{+0.6}_{-1.9}$ & 2.0$\pm$0.3 & $12^{+4}_{-3}$ 	& 9$\pm$3 & 1.4$\pm$0.2 & 79685 		& 65$\pm$20 	 \\
13~~JEM-X   & 2007 Mar 19,  01:15:03	& $1.7^{+0.3}_{-0.3}$ & $15^{+8}_{-5}$ 	& 3.6/13 		& $3.9^{+0.7}_{-2.5}$ & 3.6$\pm$0.4 & $10^{+3}_{-3}$ 	& 11$\pm$2 & 1.4$\pm$0.2 & 212715 	& 35$\pm$8 	 \\
14~~XMM  & 2007 Mar 31,  01:15:15	& $1.9^{+0.6}_{-0.4}$ & $7^{+3}_{-2}$ 	& 8.5/9 		& $1.1^{+0.2}_{-0.6}$ & 2.5$\pm$0.3 & $19^{+2}_{-2}$ 	& 21$\pm$5 & 0.6$\pm$0.1 & 1036812 	& 54$\pm$17 	 \\
15~~XMM  & 2007 Apr 02,  02:39:03		& $2.2^{+0.4}_{-0.3}$ & $6^{+2}_{-2}$ 	& 25.5/21 		& $1.3^{+0.2}_{-0.3}$ & 3.6$\pm$0.4 & $21^{+2}_{-2}$ 	& 28$\pm$5 & 0.7$\pm$0.1 & 177828 	& 54$\pm$10 	 \\
\hline
\end{tabular}
\end{tiny}
\end{minipage}
\end{table*}

\section{Discussion}
\label{sec:discussion}

Herein, we report the monitoring by \integ~and \xmm~of the Galactic Center faint transient \grs~in outburst during two visibility periods, 2005 and 2007. During these outbursts, the bolometric persistent flux of the source was observed to vary between $\sim$$0.3$ and $1.8\times10^{-9}$~\ergperseccm, with a spectral photon index always close to 2 (Table \ref{burstTable} and \ref{specpersist}). We discovered 11 type-I bursts with \integ~and also examined four other ones: two via \xmm~and two via \swift, to be as complete as possible. The parameters of the eight \rxte~bursts recently reported by \cite{galloway08} are very similar, i.e. peak fluxes, bursts fluences, rise times, and shapes, to those summarized in Table \ref{burstTable}. This argues in favor of the association proposed by these authors between the bursts and the source.
 
The distance to the source can be determined through bursts undergoing PRE. We cannot claim any of the burst detected from \grs~reached a PRE, but from the brightest one, \#4, we can calculate an upper limit on the source distance. Supposing an isotropic emission and a bolometric peak luminosity equal to the Eddington value for a pure He type-I burst (hydrogen mass fraction $X=0$), $L_{\rm Edd}=3.8\times10^{38}$~\ergpersec, as empirically derived by \cite{kuulkers03}, we obtain a distance upper limit  $d=7.6^{+2.0}_{-0.5}$ kpc. For comparison, the theoretical value \citep{lewin93} for the upper limit found by considering a He atmosphere and a canonical neutron star (1.4$M_{\odot}$ for 10 km radius), is $6.7^{+1.8}_{-0.4}$~kpc. Note also that taking a hydrogen rich mixed burst (solar composition $X=0.7$) would lead to a closer distance upper limit of $5.1^{+1.4}_{-0.3}$~kpc. These results are in agreement with the distance estimates made by \cite{cocchi99} and \cite{galloway08} and the measured high column density $\sim$$12\times10^{22}$~\cm.

Let us assume in the following that the bursts we observed were pure He bursts ($X=0$) and the source distance has a fiducial value $d=7$~kpc. Then the unabsorbed bolometric persistent flux range of the source translates to luminosities $L_{\rm pers} \approx1.7-10.5\times10^{36}$~\ergpersec. This persistent luminosity depends upon the local accretion rate per unit area, $\dot m$, via the equation $L_{\rm pers}=4\pi R^2 \dot m (GM/R)(1+z)^{-1}$, where $M$ and $R$ stand for the  mass and radius of the neutron star respectively, and $z = (1-2GM/(Rc^{2}))^{-1/2}-1$ is the gravitational redshift at its surface. We consider here the canonical values $M=1.4M_\odot$, $R=10$~km, and consequently $z=0.31$. As a results $\dot m \approx 1000-6100$~\gperseccm. A common unit for expressing $\dot m$ is the corresponding Eddington rate $\dot m_{\rm Edd}=2m_{\rm p}c(R\sigma_{\rm Th})^{-1}(1+X_0)^{-1}(1+z)$, where $m_{\rm p}$, $X_0$, $c$, and $\sigma_{\rm Th}$ are the proton mass, the H abundance of the matter accreted from the donor star, the speed of light, and the Thompson scattering cross section, respectively. From now on we will presume $X_0=0.7$ and so $\dot m_{\rm Edd}=11.5\times10^4$~ \gperseccm, leading to $\dot m \approx 0.9-5.3\%\,\dot m_{\rm Edd}$.Ê

On the other hand, the observed fluences, $f_{\rm b}$, of the bursts listed in Table \ref{burstTable}, give total energies radiated during the bursts $E_{\rm b}=4\pi d^2 f_{\rm b} \approx1.2-7.4\times 10^{39}$~erg. We can thereby estimate the ignition column of each burst, $y_{\rm ign}$, through the relation $y_{\rm ign}=E_{\rm b}(1+z)/(4\pi R^2 \epsilon_{\rm nuc})$, where $\epsilon_{\rm nuc}$ is the nuclear energy released per unit mass and which can be related to the nuclear energy released per nucleon, $Q_{\rm nuc}\approx 1.6+4X$~MeV\,nucleon$^{-1}$, by $\epsilon_{\rm nuc}=Q_{\rm nuc}\times10^{18}$~\ergperg~\citep{wallace81,fujimoto87}. This yields ignition columns $y_{\rm ign}\approx 0.8-4.8\times 10^8$~\gpercm.
    
In turn, theoretical recurrence times for the bursts, $\tau_{\rm rec}= (y_{\rm ign}/\dot m) (1+z)$, can be worked out to be~$\approx 0.2-43.9\times10^{5}$~s~$= 0.2-5.1$~days. We stress that this derivation of $\tau_{\rm rec}$ is independent of the assumed distance $d$. The observed times, $\Delta t$, between bursts (Table \ref{burstTable}) and these expected recurrence times for pure He combustion, $\tau_{\rm rec}$, match well for bursts  3 and 4 and therefore suggest they were consecutive. The same conclusion holds for bursts 1 and 2, presuming that the persistent flux in between was close to $0.3\times10^{-9}$ \ergperseccm~though we could not detect it. In order to put stronger constraints on the recurrence time, let us consider the longest uninterrupted data set, i.e. the three consecutive \xmm~revolutions of April 2007. As the theoretical recurrence time turns out to be nearly twice the measured $\Delta t$ between bursts 14 and 15, we certainly missed one burst during the data gap in the light curve shown on Fig.~\ref{lc_pn}. 
In contrast, repeating the previous calculations with a higher hydrogen abundance in the nuclear burning, $X=0.7$, return shorter recurrence times. In particular for the \xmm~bursts, $\tau_{\rm rec}$ drops to 0.3 day, which means that there should have been two bursts surrounding burst  \#15 in the light curve of revolution 1339 on Fig.~\ref{lc_pn}. Since we did not observe such bursts, we conclude that \grs~displays pure He bursts.  

Now, theoretical works predict the nature of the nuclear burning of the accreted material (H and/or He) during a type-I burst depends critically on the accretion rate of the neutron star \citep{fujimoto81,bildsten98,peng07,cooper07}. 
In short, if $0\lesssim \dot m/\dot m_{\rm Edd}\lesssim0.01$ (i), H burns unstably via the cold CNO cycle and so ignites a mixed He/H explosion. As the fusion of H involves slow $\beta$-decays, the rise time and total duration of the burst should be relatively long: $\sim$10~s and $\sim$100~s, respectively. 
For $0.01\lesssim \dot m/\dot m_{\rm Edd}\lesssim0.1$ (ii), H is burned into He steadily through the hot CNO cycle, and a pure He layer develops underneath the surface, which may ignite via the unstable 3$\alpha$ process, thereby producing a pure He burst. The strong interaction at work implies a shorter rise time and total duration ($\sim$1~s and $\sim$10~s, respectively). 
Finally, if $0.1\lesssim \dot m/\dot m_{\rm Edd}\lesssim 1$ (iii), H is not burned stably fast enough so that a pure He flash triggers a mixed He/H burst with short rise time ($\sim$1~s) and long duration ($\sim$100~s). Here, the persistent luminosity of \grs~was at the boundary of the two first cases, but as we argued above, the energetics of the bursts suggest a pure He explosion, i.e. case (ii), which is strenghtened by the fact that the observed rise times and total durations were short.

As a consistency check, let us discuss the hydrogen depletion condition. Presuming that H burns stably on the surface of the neutron star, i.e. case (ii) or (iii), at a rate fixed by $\beta$-decays $\epsilon_{\rm H}=5.8\times 10^{15}\times Z_{\rm CNO}$~erg\,s$^{-1}$\,g$^{-1}$ \citep{hoyle65} with $Z_{\rm CNO}$ the CNO mass fraction, there is a critical column depth, $y_{\rm d}$, above which H is completely depleted leaving only He for the explosion. $y_{\rm d}$ thus fixes the boundary between cases (ii) and (iii). In the steady-state, H burns stably as fast as it is accreted, giving $y_{\rm d}=X_0\dot m E_{\rm H}/\epsilon_{\rm H}$ \citep{cumming00} where $E_{\rm H}\approx 6.0\times10^{18}$~erg\,g$^{-1}$ is the energy release in the hot CNO cycle including neutrino energy losses \citep{wallace81}. 
Let us consider the most intense burst, \#4, which has $y_{\rm ign} \approx 4.8\times10^8$~\gpercm.
Even in the most extreme case of pure H accretion ($X_0=1$), we always find $y_{\rm ign} > y_{\rm d}$ as long as $Z_{\rm CNO}>0.004$. Since a CNO abundance of 0.004 for the companion star is relatively low for a population II low-mass star, we are confident that this condition on the metallicity is fullfilled and thereby that burst 4 was a pure He burst (case (ii)). The conclusion of the comparison between the depletion and ignition columns is not as straightforward for the other bursts though, as the assumed values for $X_0$ and $Z_{\rm CNO}$ turn out to be more critical. We note however that \grs~possibly lies in the central molecular zone of the GC, where \cite{najarro09} recently measured the CNO composition to be up to twice solar, $Z_{\rm CNO}=0.02$ \cite[see][for solar abundances]{asplund05}. Besides, provided the donor star is an evolved low-mass star with a He-rich composition ($X_0<0.3$), we again find $y_{\rm ign} > y_{\rm d}$ for all the bursts. 
But in view of the high uncertainties on the true values of $X_0$ and $Z_{\rm CNO}$, we stress that this depletion condition which consolidates case (ii) should only be regarded as suggestive.

\cite{kuulkers08} recently raised the issue of bursts during the quiescent state of a soft X-ray transient (SXT) and the possible role of trigger they may play for the SXT outburst itself. For \grs, the burst seen by \swift/BAT in January 2007 would be a good candidate for a trigger of the 2007 outburst. But the detection of a short activity period, prior to the long 2007 one, in September 2006 \citep{degenaar09} points to the fact that the companion star was already dumping a substantial amount of matter towards the neutron star at the time. In all likelihood, this helped the SXT outburst develop and then paved the way for type-I bursts, and not the opposite. On top of that, it is very implausible that the source produces bursts during its quiescent state at $10^{32-33}$~\ergpersec. 
This is theoretically motivated by the expected stability of hydrogen burning via the {\it pp}-process or pycnonuclear reactions below $10^{33}$~\ergpersec~\citep{fushiki87}. This is also observationally corroborated by the non-detection of type-I bursts at this luminosity level up to now (Table~1). Note that we explored 2.8~Ms of quiescent state with \integ~over four years without success.
Moreover, at $10^{33}$~\ergpersec, it would already take five years to accumulate the necessary fuel on the neutron star surface for a burst with a characteristic energy release $\sim$$10^{39}$~erg. So it looks as though \grs~gives rise to thermonuclear explosions only when accreting at a luminosity level in excess of $10^{35-36}$~\ergpersec.

While "burst-only sources" have often been put forward as prototypes for the poorly observed H burning bursts at low accretion rates, i.e. case (i), and potentially during quiescence, it seems from this discussion that \grs~is a member of the faint transient class exhibiting pure helium runaways, i.e. case (ii), only when in outburst. This is another indication that the "burst-only sources" should maybe not be held as a distinctive class. \cite{cornelisse04}, indeed, already alluded to the fact that the burst-only sources do not form a separate homogeneous group, based on the discrepancies in the bursts durations and the nature of the system (faint transient or quiescent neutron star transients).

\section{Conclusions}
Since its discovery, \grs~has not been an easy target for X-ray missions, because it is transient, faint and located in a region affected by source confusion. As a consequence, the detailed analysis of its bursting behavior needs long exposures, high resolution and sensitivity, which could finally be achieved with \integ, \xmm, and \swift. In this paper we essentially analyzed all the data on \grs~ever recorded by the two first satellites since 2002. We confirm what had been found by other instruments earlier, i.e. that the source is at a distance of $\sim$7~kpc and exhibits bursts only when in outburst, every couple of years. From the discovery of 15 new bursts in 2005/2007 and the spectral fitting of the contemporaneous persistent emission, we were able to investigate the nuclear burning processes powering the explosions. We find that unstable pure He fusion matches well the observations, whereas the presence of H would contradict the measured recurrence times and bursts durations. In addition, bursts oscillations have been proposed to be present in some bursts of \grs~by \cite{galloway08}, which is in line with the fact that these oscillations appears mostly in helium burning runaways \citep{cumming00,narayan07}. However, the  existence of a fast spinning neutron star in \grs~is still something that future work with current X-ray facilities need to conclusively demonstrate.


\begin{acknowledgements}
	GT wishes to warmly thank M. Renaud,  J. Chenevez, D. G\"otz, F. Mattana, G. Ponti, and J.A. Zurita Heras for valuable discussions and help regarding several aspects of the \integ~and \xmm~data analysis. EB is grateful to M. Perri for his help with the \swift~data analysis. MF acknowledges the French Space Agency (CNES) for financial support. 
	Part of this work has been funded by the french Agence Nationale pour la Recherche (ANR) through grant ANR--06--JC--0047.
	
        \integ~is an ESA project with instruments and science data center funded by ESA member states (especially the PI countries: Denmark, France, Germany, Italy, Switzerland, and Spain), the Czech Republic, and Poland, and with the participation of Russia and the US. 
	The \xmm~project is an ESA Science Mission with instruments and contributions directly funded by ESA Member States and the USA (NASA).     
\end{acknowledgements}


\bibliographystyle{aa}

\bibliography{gc_trap.bib}

\end{document}